\documentclass[mathleft
]{an}
\usepackage{graphicx}
\usepackage{times}
\begin{document}

\Pagespan{1}{}
\Yearpublication{\ldots}%
\Yearsubmission{\ldots}%
\Month{\ldots}%
\Volume{\ldots}%
\Issue{\ldots}%

\title{Estimating black hole masses in young radio sources using CFHT spectroscopy}

\author{M.F. Gu\inst{1}\fnmsep\thanks{Corresponding author:
  \email{gumf@shao.ac.cn}\newline}
\and S. Pak\inst{2} \and  L.C. Ho\inst{3}}
\titlerunning{Black hole masses in young radio sources}
\authorrunning{M.F. Gu, S. Pak \& L.C. Ho}
\institute{ Key Laboratory for Research in Galaxies and Cosmology,
Shanghai Astronomical Observatory, Chinese Academy of Sciences, 80
Nandan Road, Shanghai 200030,China \and Department of Astronomy and
Space Science, Kyung Hee University, Yongin-si, Gyeonggi-do 446-701,
South Korea \and The Observatories of the Carnegie Institution of
Washington, 813 Santa Barbara Street, Pasadena, CA 91101-1292}

\received{\ldots} \accepted{\ldots} \publonline{\ldots}


\abstract{The correlation between black hole masses and stellar
velocity dispersions provides an efficient method to determine the
masses of black holes in active galaxies. We obtained optical
spectra of a Compact-Steep-Spectrum (CSS) galaxy 4C +29.70, using
the Canada-France-Hawaii Telescope (CFHT) equipped with OSIS, in
August 6, 2003. Several stellar absorption features, such as Mg I
(5175\AA), Ca E band (5269\AA) and Na D (5890\AA), were detected in
the spectra. The stellar velocity dispersion, $\sigma$, of the host
galaxy, measured from absorption features is $\rm \approx 250
km~s^{-1}$. If 4C +29.70 follows the $\rm M_{BH}-\sigma$ relation
established for nearby galaxies, then its central black hole has a
mass of $\rm \approx3.3\times10^{8}M_{\odot}$. In combination with
the black hole masses of seven GPS galaxies in Snellen et al.
(2003), we find that the average black hole mass of these eight
young radio sources is smaller than that of the Bettoni et al.
(2003) sample of extended radio galaxies. This may indicate that
young radio sources are likely at the early evolutionary stage of
radio galaxies, at which the central black holes may still undergo
rapid growth. However, this needs further investigations.}


\maketitle

\section{Introduction}

Gigahertz-Peaked-Spectrum (GPS: projected linear size D$<$ 1 kpc)
and the more extended Compact-Steep-Spectrum (CSS: D$<$ 15 kpc)
sources make up $\sim$40\% of radio-loud AGNs, yet their nature is
not fully understood (see, e.g., O'Dea 1998, and references
therein).
%
%
There are two main theories to describe these compact sources: the
{\em `frustration' scenario} (van Breugel 1984), and the {\em
`youth' scenario} ( Fanti et al. 1995). In the {\em `frustration'
scenario}, the source is enshrouded by a gas and dust cocoon so
dense that its radio jets cannot escape nuclear confinement -- which
frustrates radio source growth. Alternatively, the {\em youth
scenario} attributes compactness to evolution -- if we observe it
young, the radio jet will not have expanded much, and will still be
relatively small. In this scenario, the GPS \& CSS sources represent
the earliest stages in the radio-loud AGN life cycle, before they
expand into the large-scale `classical' doubles. This is the
currently preferred theory, which is supported by estimated
dynamical ages for GPS sources of $\rm t_{dyn}\sim 10^2-10^3$ yrs
(e.g.~Owsianik, Conway \& Polatidis 1998; Tschager et al.~2000), and
by radio spectral ages for the larger CSS sources of $\rm t_{sp}<$
10$^4$ yrs (Murgia et al.~1999). These compact AGNs are then young
radio sources, either emerging for the `first time', or recently
`born again'.
%
%

There is much evidence suggesting that a significant fraction of
radio-loud AGNs exhibit radio activity which is very likely
triggered by mergers of two or more galaxies, at least one of which
is gas-rich (Heckman et al.~1986). Many compact radio source
counterparts show optical features attributed to mergers
(e.g.~double nuclei, tidal tails, arcs, and distorted isophotes),
implying that these sources are observed shortly after merging
(Stanghellini et al.~1993), before the system has settled back down.
%
%
All activity-inducing processes clearly involve injecting large
amounts of gas and dust into the nuclear regions -- at least
initially, the radio source will be in a dense and dusty
environment, with a huge reservoir of material poised to feed the
central massive black hole (MBH) at high accretion rates.
%
%

According to AGN unification models, GPS \& CSS radio sources are
expected to harbor an obscured quasar (Fanti et al.~2000). In the
past, there have been strong indications that jet power is
proportional to AGN bolometric luminosity (e.g.~Baum \& Heckman
1989; Rawlings \& Saunders 1991; Falcke et al.~1995). As GPS \& CSS
sources are powerful radio-loud AGNs, we may expect them to appear
as optically bright quasars when their jet axes are viewed close to
the line of sight. Yu \& Tremaine (2002) proposed that the growth of
high-mass MBHs comes mainly from accretion during the optically
bright QSO phases. While the `classical' radio doubles have already
been growing over long times, GPS \& CSS sources are more likely to
be in the early stages of accretion and radio-loud activity, with
the central MBH still undergoing rapid growth. Since these compact
sources are believed to evolve into the `classical' extended ones,
it is important to compare the $\rm M_{BH}$-distribution in young
radio sources with that of their larger descendants (e.g. Bettoni et
al. 2003), in order to constrain the proposed {\em `youth'
scenario}, as well as the nature of accretion in compact radio
sources. Such a comparison may also yield clues on the radio
activity in these sources, e.g. `first time' activity vs.~`born
again' scenarios, as the $\rm M_{BH}$ values should be significantly
larger in the latter case.

A key advance in estimating the central MBH masses ($\rm M_{BH}$)
within galactic bulges was the discovery of the $\rm M_{BH}-\sigma$
relation, a tight correlation between the MBH mass and the galactic
bulge stellar velocity dispersion $\sigma$ (Ferrarese \& Merritt
2000; Gebhardt et al.~2000a;
 Tremaine et al.~2002).
This relation provides an important consistency check against masses
determined from reverberation mapping of Seyfert nuclei (Nelson
2000; Gebhardt et al.~2000b; Ferrarese et al.~2001). The tightness
of the correlation over a wide range of galaxy types, both
elliptical and spiral, means that $\rm M_{BH}$ can be estimated
simply by measuring $\sigma$ of the AGN host galaxy bulge; recently,
MBH masses were estimated using this relation for a large sample of
low-$\rm z$ radio galaxies (Bettoni et al.~2003).

The optical light from GPS \& CSS radio sources, with optical
identification and known galaxy morphologies, is thought to be
dominated by old stellar populations (e.g., Snellen et al.~1999),
making them suitable for measuring the velocity dispersion,
$\sigma$, of the host galaxy. By measuring $\sigma$ from the stellar
absorption lines in the host galaxies, we may then estimate the MBH
mass using the $\rm M_{BH}-\sigma$ relation.
Here we present the optical spectrum of a CSS source 4C +29.70,
observed with the Canada-France-Hawaii Telescope (CFHT). By
measuring the central stellar velocity dispersion $\sigma$ of the
host galaxy from the stellar absorption lines, we estimated the
black hole mass using the $\rm M_{BH}-\sigma$ relation established
for nearby galaxies. 

\section{OBSERVATIONS AND REDUCTIONS}

Observations were obtained in 2003 August 6, using the 3.6m
Canada-France-Hawaii Telescope, equipped with the Optionally
Stabilized Imager and Spectrometer (OSIS). Spectra were secured
using the O600 grism to cover the spectra region 4900\AA - 7100 \AA,
at 0.57\AA~pixel$^{-1}$ dispersion. This allows us to measure the
absorption lines of Mg I (5175\AA), Ca E band (5269\AA), and Na D
(5890\AA), and other absorption line blends from the host galaxy. A
$0\farcs58$ wide slit, with a position angle of $90^{\circ}$, was
used for all the observations. Initially, we planed to observe a
sample of 27 GPS/CSS galaxies, however, we only succeeded to observe
one CSS source 4C + 29.70 due to poor weather during observing run.
The total exposure time for 4C +29.70 was 30 minutes.

The chosen grism, combined with the slit, yields a spectral
resolution for a velocity dispersion measurement of $\rm \sim 60~
km~s^{-1}$, which is adequate for the expected range of $\sigma$ in
luminous elliptical galaxies (e.g., Djorgovski \& Davis 1987;
Bender, Burstein \& Faber 1992) such as the hosts of radio galaxies.
In addition to the spectra of our source, we acquired a spectrum of
the bright star HD 224060, of type K3III, that exhibits a low
rotational velocity (Vsini$\rm <20kms^{-1}$). It is used as the
template of zero velocity dispersion.

\begin{figure}[t]
\resizebox{\hsize}{!}{\includegraphics{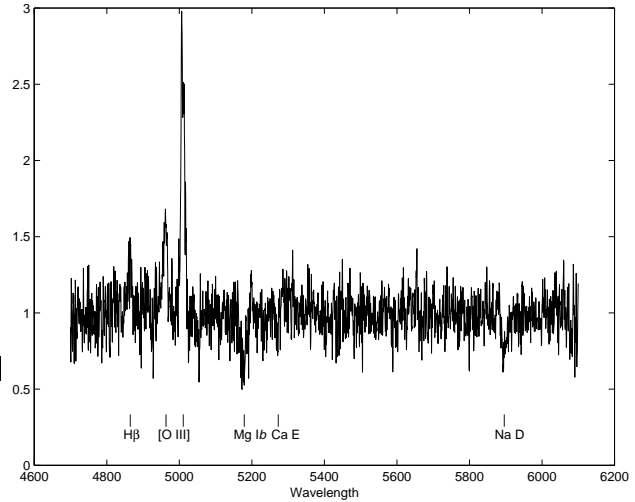}} \caption{Rest
frame spectrum of the CSS source 4C +29.70 (z=0.13069). The spectrum
has been continuum-normalized to unity. The spectrum has been
smoothed with a boxcar function of 3 pixels in order to slightly
improve the S/N for the sake of the presentation. Various absorption
features, such as Mg I (5175\AA), Ca E band (5269\AA), and Na D
(5890\AA), along with emission lines of [O III] 4959, 5007\AA, and
$\rm H\beta$, are marked. The horizontal axis is wavelength in \AA,
in the rest frame.} \label{fig1}
\end{figure}

During the observations, the seeing ranged between $0\farcs8$ and
$1\farcs0$. The targets were centered into the slit, and then the
one-dimensional spectrum was extracted from an aperture of
$2\farcs5$ diameter. Standard data reduction was applied to the
spectra using the tasks available in the {\sc IRAF}\footnote{{\sc
IRAF} is distributed by the National Optical Astronomy
Observatories,  which are operated by the Association of
Universities for Research in Astronomy,  Inc.,  under cooperative
agreement with the National Science Foundation.} package. This
procedure includes the bias subtraction, flat-fielding, wavelength
calibration using an exposure of a He-Ne-Ar lamp, and extraction of
one-dimensional spectra. We took two spectra and combined them in
order to remove cosmic-ray hits and other occasional spurious
signals in the detector. The continuum-normalized spectrum of 4C
+29.70 is displayed in the rest-frame in Fig. \ref{fig1}. Several
stellar absorption features are clearly visible in this spectrum,
which include Mg I (5175\AA), Ca E band (5269\AA), and Na D
(5890\AA). Strong emission lines of [O III] 4959, 5007 \AA~ are also
detected in 4C +29.70, but $\rm H\beta$ is only marginally detected.

\begin{figure}[t]
\resizebox{\hsize}{!}{\includegraphics{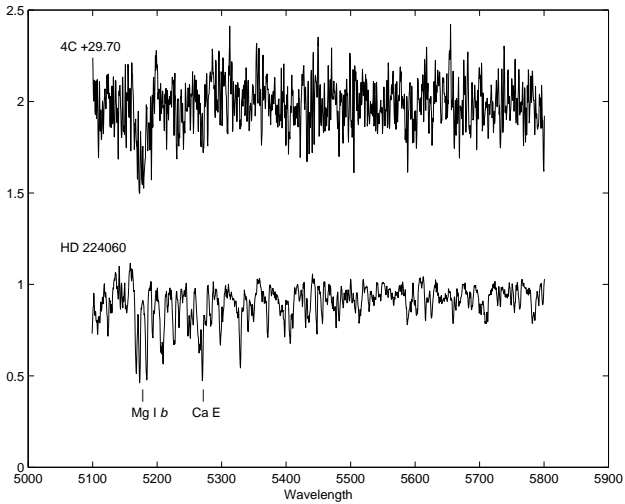}}
\caption{Spectra of 4C +29.70 (z=0.13069) and the template star HD
224060. The spectral regions are adopted in the cross-correlation
method to estimate the stellar velocity dispersion. The spectra have
been continuum-normalized to unity. The spectrum of 4C +29.70 has
been shifted vertically by 1.0 unit to avoid overlap. The object
spectrum has been smoothed with a boxcar function of 3 pixels in
order to slightly improve the S/N for the sake of the presentation.
Various absorption features, Mg I (5175\AA), and Ca E band (5269\AA)
are marked. The horizontal axis is wavelength in \AA, in the rest
frame.} \label{fig2}
\end{figure}

The stellar velocity dispersion $\sigma$ was determined using the
cross-correlation method (e.g. Tonry \& Davis 1979) implemented in
the IRAF RV package. In Fig. \ref{fig2}, the spectral regions
adopted in the cross-correlation method are presented, both in 4C
+29.70 and in the template star HD 224060. These regions comprise of
Mg I (5175\AA), Ca E band (5269\AA), and many additional absorption
blends.

In principle, in the cross-correlation method, the largest peak in
the cross-correlation function is assumed to be Gaussian shaped with
dispersion $\mu$. Also, the spectrum of the template star is assumed
to have a Fourier transform that is approximately Gaussian in
amplitude, but with unconstrained phases, and a dispersion
appropriate for a typical feature in the spectrum to have width
$\tau$. The stellar velocity dispersion of the object is then
estimated from $\mu$ in conjunction with $\tau$ (Tonry \& Davis
1979). Here we follow the same procedure used by Ho \& Filippenko
(1996a, 1996b). The cross-correlation function between the object
and the template spectrum peaks at the relative radial velocity
dispersion of the object. The correspondence between the width of
the peak and the actual velocity dispersion is determined
empirically by artificially broadening the template spectra with
Gaussians of various dispersions and by subsequently remeasuring
them using the cross-correlation algorithm. Finally, we obtained
$\rm \sigma\approx250km~s^{-1}$ for 4C +29.70.

\section{RESULTS AND DISCUSSIONS}

4C +29.70 is a CSS source of intermediate strength, selected from
the 408-MHz Bologna B2.1 catalogue (Saikia et al.\ 2002). Its
redshift is $z=0.13069$, corresponding to a distance of 476 Mpc
($H_{0}=~\rm~75~km~s^{-1}~Mpc^{-1}$ and $q_{0}=0.5$). It was
observed in snapshot mode with the Very Large Array (VLA) A-array at
4835 MHz on 1985 February 10 (Saikia et al.\ 2002). The radio
structure is highly asymmetric and double, and the total flux
density is 263 mJy (Saikia et al. 2002). The largest angular size
(LAS) of this source is 1.2 arcsec, corresponding to a projected
linear size of 2.45 kpc. The spectral index $\alpha$, between 327
MHz and 5 GHz, is about 0.7 ($\rm f_{\nu}\propto\nu^{-\alpha}$). 4C
+29.70 was optically identified as a galaxy. 

To determine $\rm M_{BH}$ from the velocity dispersion, we use the
fit of the $\rm M_{BH}-\sigma$ relation from Tremaine et al. (2002),
which is given by:
\begin{equation}
\rm log~(\frac{\it M_{\rm BH}}{\it
M_{\odot}})=(8.13\pm0.06)+(4.02\pm0.32)~log~(\frac{\sigma}{200~km
~s^{-1}}).
\end{equation}
This relation was derived using a carefully culled sample of the
best-quality black hole mass measurements, and a fitting technique
that properly accounts for uncertainties in both $\rm M_{BH}$ and
$\sigma$.


The measurements used to calibrate the above relation were the
luminosity-weighted velocity dispersion $\sigma_{e}$, measured in a
slit aperture of length 2$r_{e}$. However, our value of $\sigma$
does not exactly correspond to this aperture size. Velocity
dispersions can be corrected to a standard aperture size using the
relations given by J$\o$rgensen, Franx \& Kj$\ae$rgaard (1995),
but
the corrections depend on the effective radii, which are not known
accurately for 4C +29.70. Rather than apply an uncertain correction
to the measured velocity dispersion, we choose to use our measured
dispersion directly to estimate $\rm M_{BH}$. As shown by Gebhardt
et al. (2000a), the value of $\sigma$ measured from slit apertures
of different length differ from $\sigma_{e}$ by less than 10\% for
typical, nearby elliptical galaxies.

Using the result of $\rm \sigma\approx250 km~s^{-1}$ in conjunction
with the $\rm M_{BH}-\sigma$ relation, we estimate a black hole mass
of $\rm \approx3.3\times10^{8}M_{\odot}$ for 4C +29.70. Recently,
MBH masses were estimated using either the $\rm M_{BH}-\sigma$
relation, or the black hole mass - bulge luminosity relation, for a
large sample of low-$\rm z$ radio galaxies (Bettoni et al.~2003).
They found the black hole masses to range from $~5\times 10^{7}$ to
$\rm ~6\times 10^{9}~ M_{\odot}$, with an average $\rm \langle
log(\it M_{\rm BH}/\it M_{\odot})\rm \rangle\sim8.9\pm0.4$. To
compare the black hole masses of young radio sources to those of
extended radio galaxies, we tentatively combine the black hole
masses of seven GPS galaxies estimated in Snellen et al. (2003) with
our measurements. The average value of black hole masses of these
eight young radio sources is $\rm \langle log(\it M_{\rm BH}/\it
M_{\odot})\rm \rangle=8.1\pm0.4$. This value is apparently smaller
than that of Bettoni sample. This may indicate that the young radio
sources are likely at the early evolutionary stage of radio
galaxies, at which the central black holes may still undergo rapid
growth. To evolve into the extended radio galaxies in $\sim10^7 -
10^8$ yr, averagely they must accrete at near (or even super-)
Eddington luminosity, i.e. $\sim3\rm M_{\odot}yr^{-1}$ for a
$10^{8.1}\rm M_{\odot}$ black hole, from which the star formation
rate of several hundred $\rm M_{\odot}yr^{-1}$ is roughly expected
(Wu \& Cao 2006). According to the bolometric correction of Hopkins
et al. (2007), the hard X-ray luminosity of $\sim 10^{44} \rm erg
~s^{-1}$ is then expected, which is consistent with the typical
value of GPS galaxies (e.g. Guainazzi et al. 2006). However, the
expected MFIR luminosity $\sim 10^{45} \rm erg ~s^{-1}$ (Hopkins et
al. 2007; Satyapal et al. 2005) are somehow larger than the
observations (e.g. Heckman et al. 1994; Fanti et al. 2000).
Nevertheless, our conclusion can not be firmly drawn at present
stage due to the small sample size.
Alternatively, young radio source can be `born again' sources.The
detection of quasar remnants in nearby galaxies, in the form of
inactive massive BHs, conclusively demonstrates that the cosmic
evolution of AGN activity must be episodic (Richstone et al. 1998).
The duty cycle for accretion may be quite short, so that any
individual massive BH is likely to have been activated (and
deactivated) many times since it was formed. If this scenario is
true, young radio sources may host black holes with masses
comparable to those of extended radio galaxies.


Apart from the masses of central black hole, the nature of accretion
in young radio sources can help us understand the evolution of radio
galaxies. For a given supermassive black hole, the full life cycle
of a radio galaxy is believed to be: GPS $\longrightarrow$ CSS
$\longrightarrow$ FR II $\longrightarrow$ FR I (Marecki et al.
2003). In the unification scheme, FR II and FR I radio galaxies are
unified with FSRQs and BL Lacs, respectively (Urry \& Padovani
1995). For the population of blazars (i.e. FSRQs and BL Lacs), their
accretion rate is believed to decrease along the evolutionary
sequence FSRQs $\longrightarrow$ BL Lacs (Wang, Ho \& Staubert
2003), in which FSRQs are rich in gas and therefore are
characterized by large accretion rates, while BL Lac objects
represent evolved sources depleted of gas, with faint nuclear
emission and low-power jets (Cavaliere \& D'Elia 2002; B\"{o}ttcher
\& Dermer 2002). When viewed at large angles to our line of sight,
the sequence becomes: FR II $\longrightarrow$ FR I, where FR II
sources have higher accretion rates compared to FR I ones.
Ghisellini \& Celotti (2001) recently suggested that the dividing
line between FR II and FR I corresponds to a transition in the
accretion mode, from a SS disk (FR II) to an optically thin ADAF (FR
I) one. If we fit GPS/CSS sources into this evolutionary sequence,
their accretion rates are then expected to be relatively high.



In conclusion, we estimated the black hole mass for 4C +29.70, a CSS
source, as a first step of our project to constrain the evolutionary
scenario of radio galaxies through systematically comparing the
black hole masses between young radio sources and extended radio
galaxies. In combination with the black hole masses of seven GPS
galaxies, we find that the average black hole mass of young radio
sources are smaller than that of extended radio galaxies. This may
indicate that the young radio sources are likely at the early
evolutionary stage of radio galaxies, at which the central black
holes may still undergo rapid growth. However, this result should be
re-investigated with a large and complete sample, of which the
comparison with the well constructed sample of extended radio
galaxies is required as well. This is our ongoing project.

\acknowledgements

This work is supported by NSFC (grants 10633010, 10703009, 10821302
and 10833002) and by 973 Program (No. 2009CB824800).

\newpage


\begin{thebibliography}{}

\bibitem{}Baum, S.A., Heckman, T.: 1989, ApJ~336, 702
\bibitem{}Bender, R., Burstein, D., Faber, S.M.: 1992, ApJ~399, 462
\bibitem{}Bettoni, D., Falomo, R., Fasano, F., Govoni, F.: 2003, A\&A~399, 869
\bibitem{}B\"{o}ttcher, M., Dermer, C.D.: 2002, ApJ~564, 86
\bibitem{}Cavaliere, A., D'Elia, V.: 2002, ApJ~571, 226
\bibitem{}Djorgovski, S., Davis, M.: 1987, ApJ~313, 59
\bibitem{}Falcke, H., Malkan, M.A., Biermann, P.L.: 1995, A\&A~298, 375
\bibitem{}Fanti, C., Fanti, R., Schilizzi, R.T., Spencer, R.E., Stanghellini, C.: 1995, A\&A~302, 317
\bibitem{}Fanti, C., Pozzi, F., Fanti, R., et al.: 2000, A\&A~358, 499
\bibitem{}Ferrarese, L., Merritt, D.: 2000, ApJ~539, L9
\bibitem{}Ferrarese, L., Pogge, R.W., Peterson, B.M., Merritt, D., Wandel, A., Joseph, C.L.: 2001, ApJ~555, L79
\bibitem{}Gebhardt, K., Bender, R., Bower, G., et al.: 2000a, ApJ~539, L13
\bibitem{}Gebhardt, K., Kormendy, J., Ho, L.C., et al.: 2000b, ApJ~543, L5
\bibitem{}Ghisellini, G., Celotti, A.: 2001, A\&A~379, L1
\bibitem{}Guainazzi, M., et al.: 2006, A\&A~446, 87
\bibitem{}Heckman, T.M., et al.: 1986, ApJ~311, 526
\bibitem{}Heckman, T.M., et al.: 1994, ApJ~428, 65
\bibitem{}Ho, L.C., Filippenko, A.V.: 1996a, ApJ~466, L83
\bibitem{}Ho, L.C., Filippenko, A.V.: 1996b, ApJ~472, 600
\bibitem{}Hopkins, P.F., et al.: 2007, ApJ~654, 731
\bibitem{}J$\o$rgensen, I., Franx, M., Kj$\ae$rgaard, P.: 1995, MNRAS~276, 1341
\bibitem{}Marecki, A., Spencer, R.E., Kunert, M.: 2003, PASA~20, 46
\bibitem{}Murgia, M., Fanti, C., Fanti, R., Gregorini, L., Klein, U., Mack, K.-H., Vigotti, M.: 1999, A\&A~345, 769
\bibitem{}Nelson, C.H.: 2000, ApJ~544, L91
\bibitem{}O'Dea, C.P.: 1998, PASP~110, 493
\bibitem{}Owsianik, I., Conway, J.E., Polatidis, A.G.: 1998, A\&A~336, L37
\bibitem{}Rawlings, S., Saunders, R.: 1991, Nature~349, 438
\bibitem{}Richstone, D.O., Ajhar, E.A., Bender, R., et al.: 1998, Nature~395, A14
\bibitem{}Saikia, D.J., Thomasson, P., Spencer, R.E., Mantovani, F., Salter, C.J., Jeyakumar, S.: 2002, A\&A~391, 149
\bibitem{}Satyapal, S., et al.: 2005, ApJ~633, 86
\bibitem{}Snellen, I.A.G., Schilizzi, R.T., Bremer, M.N., Miley, G.K., de Bruyn, A.G., R\"{o}ttgering, H.J.A.: 1999, MNRAS~307, 149
\bibitem{}Snellen, I.A.G., Lehnert, M.D., Bremer, M.N., Schilizzi,
R.T.: 2003, MNRAS~342, 889
\bibitem{}Stanghellini, C., O'Dea, C.P., Baum, S.A., Laurikainen, E.: 1993, ApJS~88, 1
\bibitem{}Tonry, J., Davis, M.: 1979, AJ~84, 1511
\bibitem{}Tremaine, S., Gebhardt, K., Bender, R., et al.: 2002, ApJ~574, 740
\bibitem{}Tschager, W., Schillizi, R., R\"{o}ttgering, H.J.A., Snellen, I.A.G., Miley, G.K.: 2000, A\&A~360, 887
\bibitem{}Urry, C.M., Padovani, P.: 1995, PASP~107, 803
\bibitem{}van Breugel, W.J.M.: 1984, IAU Symp.~110, 59
\bibitem{}Wang, J.M., Ho, L.C., Staubert, R.: 2003, A\&A~409, 887
\bibitem{}Wu, Q.W., Cao, X.W.: 2006, PASP~118, 1098
\bibitem{}Yu, Q.J., Tremaine, S.: 2002, MNRAS~335, 965

\end{thebibliography}
\end{document}